\documentclass[letterpaper, 10 pt, conference]{ieeeconf}  

\usepackage{amsmath}
\usepackage{amssymb}
\usepackage{amsfonts}
\usepackage{epsfig}
\usepackage{bm}
\usepackage{graphicx}
\usepackage{caption}
\usepackage{subcaption}
\usepackage{float}
\usepackage{algorithm}
\usepackage{algorithmicx}
\usepackage{algpseudocode}
\usepackage{soul}

\newcommand{\bmtx}{\begin{bmatrix}}
\newcommand{\emtx}{\end{bmatrix}}
\newcommand{\bsmtx}{\left[ \begin{smallmatrix}} 
\newcommand{\esmtx}{\end{smallmatrix} \right]}
\newcommand{\bmatarray}[1]{\left[\begin{array}{#1}}
\newcommand{\ematarray}{\end{array}\right]} 
\newcommand{\field}[1]{\mathbb{#1}}
\newcommand{\R}{\field{R}}

\title{\LARGE \bf
Recovering Robustness in Model-Free Reinforcement Learning
}

\author{Harish K Venkataraman$^{1}$ and Peter J.
  Seiler$^{2}$
  \thanks{*Supported by NSF grant NRI: Collaborative Research:
    Autonomous Quadrotors for 3D Modeling and Inspection of Outdoor
    Infrastructure}
  \thanks{$^{1}$Harish K.Venkataraman graduate student, Aerospace
    Engineering and Mechanics,University of Minnesota Twin Cities,
    Minneapolis MN 55455 {\tt\small albert.author@papercept.net}}%
  \thanks{$^{2}$Peter J. Seiler is Associate Professor in Aerospace
    Engineering and Mechanics,University of Minnesota Twin Cities,
    Minneapolis MN 55455 {\tt\small b.d.researcher@ieee.org}}%
}

\begin{document}

\maketitle
\thispagestyle{empty}
\pagestyle{empty}

\begin{abstract}

  Reinforcement learning (RL) is used to directly design a control
  policy using data collected from the system.  This paper considers
  the robustness of controllers trained via model-free RL. The
  discussion focuses on posing the (model-free) linear quadratic
  Gaussian (LQG) problem as a special instance of RL.  A simple LQG
  example is used to demonstrate that RL with partial observations can
  lead to poor robustness margins. It is proposed to recover
  robustness by introducing random perturbations at the system input
  during the RL training.  The perturbation magnitude can be used to
  trade off performance for increased robustness.  Two simple examples
  are presented to demonstrate the proposed method for enhancing
  robustness during RL training.

\end{abstract}

\section{INTRODUCTION}

There has been rapid and impressive progress in machine learning in
the past decade.  One particular approach, reinforcement learning (RL)
\cite{sutton98,szepesvari10}, has close connections to optimal control
techniques.  RL is a model-free approach to directly design the
control policy using data collected from the system via simulation or
experiments. There have been several successful applications of RL on
a variety of systems including helicopters \cite{ng04} and robotics
\cite{peters06,stulp10,gudimella17,kalashnikov18}.

This paper uses the standard model-based linear quadratic Gaussian
(LQG) problem to explore the robustness of model-free RL controllers.
The LQG problem, reviewed in Section~\ref{subsec:lqg}, is formulated
with linear state-space models and an expected quadratic cost
\cite{kwak72,zhou95}.  The optimal controller is an
observer/state-feedback with gains computed from two Riccati
equations.  We refer to this as ``model-based'' because the optimal
controller is constructed explicitly using the state matrices. RL,
reviewed in Section~\ref{subsec:rl}, is a closely related problem
formulated with partially observable Markov decision processes
(POMDPs) and expected cumulative rewards \cite{sutton98,szepesvari10}.
We refer to this as ``model-free'' because typical solution methods
directly search for the control policy using simulation or
experimental data.\footnote{The boundaries between ``model-based'' and
  ``model-free'' are not necessarily well-defined.  For example, the
  state matrices used in the LQG problem are often constructed from
  data using system identification techniques. We still call the LQG
  solution ``model-based'' as the construction of the optimal
  controller directly uses these state matrices. The sample complexity
  of such an approach is a subject of current research \cite{dean17}.
  Conversely, an RL policy trained with simulation data implies the
  existence of a model, i.e. the simulator itself.  We still call
  this ``model-free'' as the RL controller is constructed from data
  without directly using the model information.}

RL with POMDPs is sufficiently general to solve the LQG problem as a
special case, as discussed in Section~\ref{subsec:solvinglqg}.  This
connection is motivated by recent work \cite{recht18} which considers
the linear quadratic regulator (LQR) as a special instance of RL with
MDPs.  Section~\ref{sec:RLrobust} builds on \cite{recht18} by
exploring the robustness properties of model-free RL with
POMDPs. First, Section~\ref{subsec:doyleLQG} reviews a well-known
example by Doyle~\cite{doyle78} in which the optimal LQG controller
has poor robustness margins. This is in contrast to LQR state feedback
controllers which have provably good margins \cite{anderson07}.
Section~\ref{subsec:doyleRL} finds (nearly) optimal policies for
Doyle's example using a simple RL method: policy search using gradient
ascent and random initialization.  With sufficient data, this
policy search converges to the optimal LQG controller.
This illustrates that model-free RL with POMDPs, as a special case of
LQG, can also lead to controllers with poor robustness margins.

Small robustness margins indicate that the feedback system may become
unstable due to small changes in the plant gain or parasitic
dynamics. This has practical implications for model-free RL with
POMDPs.  Small robustness margins imply that an RL controller trained
via simulation might lead to an unstable feedback system when
implemented on the real plant.  Alternatively, consider the scenario
where the RL controller is trained via experimental data on a real
physical device. The RL controller might cause instability if the
dynamics of the system vary slightly over time.  Moreover, the same RL
controller might cause instability if implemented on other identical
plants due to manufacturing tolerances, e.g. RL trained on one robot
but implemented for production on many of the same type of robot.

Several methods were proposed to recover robustness in LQG regulators
including loop transfer recovery \cite{doyle79,doyle81} and robust
$H_2$ \cite{paganini95}. These issues also motivated the development
of alternative synthesis and analysis techniques including $H_\infty$
optimal control \cite{dgkf89}, $\mu$ analysis
\cite{doyle82,packard93mu}, and DK synthesis \cite{packard93dk}. All
these approaches to address robustness issues can be characterized as
model-based.

A key contribution of this paper is a model-free method to enhance
robustness of RL controllers. This approach, discussed in
Section~\ref{subsec:recoverRL}, consists of introducing random
perturbations at the system input during the RL training phase.  The
specified level of input perturbation provides a tuning knob to
trade-off performance for robustness.  It is shown that this
modification to RL training improves the robustness margins on Doyle's
example (Section~\ref{subsec:recoverRL}) and a simplified model of a
flexible body (Section~\ref{sec:application}).

\section{Optimal Control Formulations and Solutions}


\subsection{Linear Quadratic Gaussian (LQG) Control}
\label{subsec:lqg}
This section briefly reviews the LQG control problem and its
solution. Additional details on LQG and the more general $H_2$
optimal control problem can be found in many textbooks,
e.g. Chapter 6 of \cite{kwak72} and Chapter 14 of \cite{zhou95}.

Consider a linear time-invariant, discrete-time system:
\begin{align}
\label{eq:LTIDT}
\begin{split}
x_{t+1} & = A x_t + B u_t + B_w w_t \\
y_t & = C x_t + v_t
\end{split}
\end{align}
where $x\in \R^{n_x}$ is the state, $u \in \R^{n_u}$ is the control
input, and $y\in \R^{n_y}$ is the measurement. The process noise
$w \in \R^{n_w}$ and sensor noise $v\in \R^{n_v}$ are assumed to be
white, zero mean, and Gaussian with variances $W:=E[w_t w_t^T]$ and
$V:=E[v_t v_t^T]$.  The (infinite-horizon) LQG optimal control problem
is formulated using a quadratic cost:
\begin{align}
  J_{LQG}(u) := \lim_{N\rightarrow \infty} \frac{1}{N} 
       E \left[ \sum_{t=0}^N x_t^T Q x_t + u_t^T R u_t \right]
\end{align}
$Q\succeq 0$ and $R\succ 0$ are matrices that define penalties on the
state and control input.  The control inputs $u_t$ are restricted to
depend on past measurements,
i.e. $u_t(y_0,\ldots,y_{t-1})$.\footnote{This form assumes a one
  time-step delay from measurements at time $t-1$ to the use for
  control at time $t$. This ``delayed'' form for LQG
  accounts for any processing and sensing delays in the feedback path.
  An alternative formulation assumes that $u_t$ depends on
  measurements up to time $t$, i.e.  $u(y_0,\ldots,y_t)$. This
  ``current'' form for LQG allows (immediate) direct feedthrough
  in the control.  This paper uses the ``delayed'' form but
  similar results can be obtained for the ``current'' form.}  The
infinite-horizon LQG problem is to select these control inputs to
minimize the cost $J_{LQG}$.


The infinite-horizon LQG problem includes additional
technical assumptions, e.g. stabilizability of $A$ and $B$. These
additional assumptions ensure that an optimal solution exists
and is given by the following estimator and state-feedback:
\begin{align}
\label{eq:lqg}
\begin{split}
\hat{x}_{t+1} & = A \hat{x}_t + B u_t 
           + L \left( y_t - C \hat{x}_t  \right) \\
u_t & = -K \hat{x}_t
\end{split}
\end{align}
The matrices $K$ and $L$ are the optimal linear quadratic regulator
(LQR) and Kalman filter gains. To compute these gains, let
$DARE(A,B,Q,R)$ denote the discrete-time Algebraic Riccati Equation
involving the matrix $X=X^T$:
\begin{align}
  X = A^{T} X A - A^{T} X B(R+B^{T} X B)^{-1} B^{T} X A + Q
\end{align}
Let $P_c$ and $P_e$ denote the stabilizing solutions to
$DARE(A,B,Q,R)$ and $DARE(A^T,C^T,B_wWB_w^T,V)$, respectively. The LQR
and Kalman filter gains are:
\begin{align}
    K & :=(B^T P_c B + R)^{-1} B^T P_c A \\
    L & := A P_e C^T(C P_e C^T + V)^{-1}  
\end{align}
The optimal controller in Equation~\ref{eq:lqg} exhibits the
well-known separation principle: it consists of the optimal
state-feedback gain coupled with an optimal state estimate.  This is a
model-based solution, i.e. a model of the plant dynamics (as given by
$A$, $B$, $C$, etc.)  is used to compute the gains and
construct the controller in Equation~\ref{eq:lqg}.  This is in
contrast to the standard approaches for reinforcement learning which
are data-driven.

\subsection{Reinforcement Learning (RL)}
\label{subsec:rl}

This section briefly reviews reinforcement learning (RL). Some
notation is chosen to align more closely with the LQG problem
discussed in the previous section.  Additional details on RL can be
found in \cite{sutton98} and \cite{szepesvari10}.

RL is used to design policies (controllers) for an agent interacting
with its environment (system/plant). Most RL problems are formulated
with a Markov Decision Process (MDP) which obey the Markovian state
assumption: the current state along with future actions completely
determine the future states. It is further assumed that the state is
available for the agent (full state feedback). It will be useful to
instead consider models that include observations with uncertainty.
These are known as Partially Observable Markov Decision Processes
(POMDPs) and are defined by:
\begin{itemize}
\item A set of states $S$,
\item A set of actions $A$,
\item A state transition probability, 
   $\mathcal{T}$ 
\item A reward function, $r: S \times A \rightarrow \R$,
\item A set of observations $O$, and
\item An observation probability, $\mathcal{O}$.
\end{itemize}
A POMDP models the interaction of an agent with its environment
accommodating both process and measurement uncertainty. The
environment at time $t$ is in a state $s_t \in S$.  The agent takes an
action $a_t \in A$ and, as a result, the environment transitions to a
new state $s_{t+1}$ with probability
$\mathcal{T}(s_{t+1} \, \vert \, s_t, a_t)$. This also generates a
reward $r(s_t,a_t)$.  Moreover, the agent receives an observation
$o_t \in O$ with probability $\mathcal{O}( o_t \, \vert \, s_t, a_t)$.
The state transition and observation probabilities capture random
variations due to environmental disturbances and measurement noise
respectively.  The objective in (finite-horizon) RL is to select the
sequence of actions to maximize the following expected cumulative
reward:
\begin{align}
\label{eq:Jrl}
  J_{RL}(a) := E\left[ \sum_{t=0}^N r(s_t,a_t) \right]
\end{align}

For MDPs the agent is assumed to have access to the full state at each
time, i.e. $o_t = s_t$. In this case, the actions can be computed by
policies $\pi : S \rightarrow A$ that map the state $s_t$ to an action
$a_t = \pi(s_t)$. This represents a deterministic policy but
stochastic stationary policies can also be used.  Standard RL
techniques compute the policy using (simulated or experimental)
data. There are a number of methods to construct policies that
maximize the cumulative reward including value iteration, policy
iteration, policy search, etc. These approaches are model free,
i.e. they require no explicit knowledge of the distribution
$\mathcal{T}$.

%

In the more general POMDP formulation, the agent only has access to
observations $o_t$. These observations provide information on the
state and action $(s_t,a_t)$ based on the probability
$\mathcal{O}$. The action $a_t$ at time $t$ is restricted to depend on
past observations and actions,
i.e. $a_t(o_0,\ldots,o_{t-1},a_0,\ldots,a_{t-1})$.  Many solution
methods exist for RL with POMDPs. In some cases, they require the
construction of a belief state (or estimate of the hidden state) from
the observations.



\subsection{Solving LQG as a Special Case of RL}
\label{subsec:solvinglqg}
The summary of RL in the previous subsection focused on finite-state
POMDPs with a finite-horizon cumulative reward.  This formulation,
with a few minor extensions, is sufficiently general to solve the LQG
problem as a special case.  This reformulation is motivated by
\cite{recht18} which solves for an linear quadratic regulator (LQR)
state feedback as a special case of RL with MDPs.

First, the LQG dynamics can be modeled as a POMDP with state, action
(control input), and observation (measurement) at time $t$ given by
$x_t$, $u_t$, and $y_t$.  This requires continuous sets for these
quantities: $S:=\R^{n_x}$, $A:=\R^{n_u}$, and $O:=\R^{n_y}$.  Thus the
transition and observation probabilities $\mathcal{T}$ and
$\mathcal{O}$ are given as probability density functions.
Specifically, the LQG plant update (Equation~\ref{eq:LTIDT}) implies
that the transition to state $x_{t+1}$ given $(x_t,u_t)$ is modeled by
a Gaussian distribution with mean $A x_t + B u_t$ and variance
$B_w W B_w^T$. Thus
$\mathcal{T} \sim \mathcal{N}( Ax_t + Bu_t, B_w W B_w^T)$
\footnote{Here $\mathcal{T}$ is the probability density of $x_{t+1}$
  given $(x_t,u_t)$. $\mathcal{T} \sim \mathcal{N}(m,\Sigma)$ denotes
  that $\mathcal{T}$ is given by the probability density function for
  a normal distribution with mean $m$ and variance
  $\Sigma$.}. Similarly, the LQG measurement $y_t$ given $(x_t,u_t)$
is modeled by $\mathcal{O} \sim \mathcal{N}( C x_t, V)$.
The per timestep RL reward corresponding to the LQG problem is:
\begin{align}
  \label{eq:lqgreward}
  r_{LQG}(x_t,u_t) := -\left( x_t^T Q x_t + u_t^T R u_t \right)
\end{align}
This is simply the negative of the per timestep LQG cost.
Section~\ref{subsec:rl} described RL with a finite horizon cumulative
reward (Equation~\ref{eq:Jrl}).  A discount factor can be introduced
to ensure that the cumulative reward remains bounded as
$N\rightarrow \infty$. Alternatively, the cost can be normalized by
$\frac{1}{N}$. Normalization is used here to align with the
infinite-horizon LQG problem. LQG, recast in the RL framework,
corresponds to maximizing the following average reward:
\begin{align}
  J_{LQG}(u) := \lim_{N\rightarrow \infty} \frac{1}{N} 
              E\left[ \sum_{t=0}^N r_{LQG}(x_t,u_t) \right]
\end{align}
Section~\ref{subsec:lqg} summarizes the typical model-based LQG
solution.  As noted above, this standard approach requires specific
knowledge of the model dynamics $(A,B,C,$ etc$)$. This approach should
be used if such model data is available since it provides the optimal
controller from simple linear algebra calculations.  Alternatively,
the LQG problem can be formulated, as discussed here, as a special
case of RL with POMDPs.  This allows for existing RL techniques to be
used to compute model-free solutions to the LQG problem. 



\section{Robustness of RL Controllers}
\label{sec:RLrobust}

\subsection{LQG Robustness Issues: Doyle's Example}
\label{subsec:doyleLQG}

This section reviews a well-known example by Doyle~\cite{doyle78} to
illustrate the robustness issues that can arise with LQG control.
Consider the discrete-time LQG problem formulated in
Section~\ref{subsec:lqg} with the following plant, noise, and cost
data:
\begin{align*}
& A := \bmtx 1.1052 & 0.1105\\ 0& 1.1052 \emtx, 
B := \bmtx 0.0053 \\ 0.1052 \emtx, 
B_w := \bmtx 0.1105 \\ 0.1052 \emtx \\
& C^T := \bmtx 1 \\ 0 \emtx,
Q := 10^3 \bmtx 1 & 1 \\ 1 &  1 \emtx,  
R := 1, W:=10^3, V := 1
\end{align*}
This corresponds to a discretization of the continuous-time plant
dynamics given in \cite{doyle78} with zero-order hold and sample
time $T_s=0.1$sec.  The optimal controller is the estimator and state
feedback in Equation~\ref{eq:lqg} with the gains:
\begin{align}
   K^T =  \bmtx 9.5193 \\ 10.2579 \emtx \mbox{ and }
   L = \bmtx 1.1297 \\ 1.0012 \emtx
\end{align}
This achieves the optimal cost $J_{LQG} = 1.373 \times 10^5$.


The feedback system of the plant and LQG controller has classical gain
margins of $[0.9802,1.0007]$.  Thus very small changes in the plant
gain will cause instability. The feedback system also has very small
phase margins of $\pm 0.070degs$. Thus any parasitic (unmodeled)
dynamics will also cause instability.  Finally, the symmetric disk
margin $m_d$ \cite{bates02,blight94} is another useful robustness
indicator\footnote{The margin $m_d>1$ defines a disk in the complex
  plane with diameter on the real axis $[\frac{1}{m_d}, m_d]$. The
  feedback system is stable for all gain and phase variations within
  this disk.}. The symmetric disk margin for this example is
$m_d=1.0007$. This is consistent with the poor classical gain and
phase margins.

A key point of Doyle's example is that LQG regulators can have
arbitrarily small margins.  This is in contrast to LQR state feedback
controllers which have provably good margins \cite{anderson07}.  The
plant in Doyle's example is unstable with both eigenvalues at $z=1.1052$.
However, poor robustness margins can arise even if the plant is stable
and minimum phase, e.g. as in \cite{moore81}.  See Section 8.3 of
\cite{anderson07} for additional details on loss of robustness with
observers.  Methods to recover robustness in LQG regulators include
loop transfer recovery \cite{doyle79,doyle81} and robust $H_2$
\cite{paganini95}. Robustness issues also motivated the development of
alternative synthesis and analysis techniques including $H_\infty$
optimal control \cite{dgkf89}, $\mu$ analysis
\cite{doyle82,packard93mu}, and DK synthesis \cite{packard93dk}. All
these approaches can be characterized as model-based.

\subsection{Solving Doyle's Example with RL}
\label{subsec:doyleRL}

This section reconsiders Doyle's example within the RL framework.  The
environment (system) is modeled by the POMDP corresponding to the
discretized dynamics from Doyle's example.  It is assumed that the
environment model is not directly available and the policy
(controller) is constructed using only input-output data.  This
corresponds to the situation where data is collected either from
simulations or from experiments.  The reward is defined as in
Equation~\ref{eq:lqgreward} with the matrices $Q$ and $R$ given in the
previous section.

The optimal LQG controller (Equation~\ref{eq:lqg}) has an observer /
state-feedback structure with explicit dependence on the model
data. In the RL framework, the policy should be
parameterized without specific dependence on the model.  The policy
for Doyle's example will be parameterized as a second-order, output
feedback system in companion form:
\begin{align}
  \label{eq:lqgcanon}
  \begin{split}
    z_{t+1} & = A_K(\theta) z_t + B_K(\theta) y_t \\
    u_t & = C_K(\theta) z_t
  \end{split}
\end{align}
where 
\begin{align}
  A_K(\theta) := \bmtx 0 & \theta_1 \\ 1 & \theta_2 \emtx, 
  B_K(\theta) := \bmtx 1 \\ 0 \emtx,
  C_K^T(\theta) := \bmtx \theta_3 \\ \theta_4 \emtx                      
\end{align}
Each vector $\theta \in \R^4$ corresponds to a specific policy denoted
by $K(\theta)$. Note, for later comparison, that the the optimal LQG
controller can be written in this companion form (via a state
transformation) as:
\begin{align}
  \theta_{LQG} := \bmtx -0.1095 & -0.0491 & -21.02 & 23.21 \emtx^T
\end{align}


Algorithm~\ref{alg:GDrandsearch} provides a method to maximize the
expected cumulative reward via gradient ascent with random
initialization. Define the search hypercube $\mathcal{H} \subset \R^4$
by
$\mathcal{H}(\underline{\theta},\bar{\theta}) := \left\{ \theta \in
  \R^4 \, : \, \underline{\theta}_k \le \theta_k \le \bar{\theta}_k,
  \,\, k=1,\ldots,4 \right\}$.  The algorithm randomly (uniformly)
samples this hypercube for an initial vector of policy parameters,
$\theta$. A gradient step is taken to move to a new parameter vector
with higher expected cumulative reward.  For simplicity, the algorithm
uses a fixed number of gradient steps $N_{ga}$. The parameter vector
is allowed to exit the initial hypercube $\mathcal{H}$ during the
gradient ascent.  This entire process is repeated for $N_{ri}$ random
initializations.  The (sub-)optimal parameter vector $\theta_{opt}$
and largest reward $J_{opt}$ are returned. As noted previously, there
are many alternative algorithms for RL with POMDPs.  Gradient ascent
with random initialization is used here because the simple
implementation allows to focus on the robustness issues.

Steps 9 and 6 in Algorithm~\ref{alg:GDrandsearch} compute the the
reward $J_{RL}$ and its gradient $\nabla J_{RL}$. A typical
implementation would estimate these values with sample means obtained
from many simulations of the closed-loop environment with the policy
$K(\theta)$.  These estimates converge to their true values as the
number of simulations tends to infinity.  This step was simplified in
our implementation of Algorithm~\ref{alg:GDrandsearch} to allow for
efficient studies with a large number of initializations.
Specifically, the true expected reward and gradient were exactly
computed in Steps 9 and 6 from the solutions of related Lyapunov
equations. Details are given in the appendix.  This abstraction
of a true RL implementation avoids the need for running
many simulations for each sample $\theta$.

%

\begin{algorithm}
\caption{Gradient Ascent with Random Initialization}
\label{alg:GDrandsearch}
\begin{algorithmic}[1]
\State \textbf{Given:} Hypercube $\mathcal{H}(\underline{\theta},\bar{\theta})$, number of initializations $N_{ri}$, gradient steps $N_{ga}$, and gradient
stepsize $\eta$. 
\State \textbf{Initialize:} $J_{opt} = -\infty$, $\theta_{opt} = 0$
\For{$i=1, \dots, N_{ri}$}
\State Sample $\theta \in \mathcal{H}$ 
  from random uniform distribution 
\For{$j=1, \dots, N_{ga}$}
\State Evaluate gradient $\nabla J_{RL}(\theta)$ 
\State Update $\theta \leftarrow \theta + \eta \nabla J_{RL}(\theta)$
\EndFor
\State Evaluate reward $J_{RL}(\theta)$ obtained by new $K(\theta)$
\If {$J_{RL}(\theta) > J_{opt}$} 
\State Set $J_{opt} := J_{RL}(\theta)$ and $\theta_{opt} := \theta$
\EndIf
\EndFor
\State \textbf{Return} $J_{opt}$ and $\theta_{opt}$ 
\end{algorithmic}
\end{algorithm}

Doyle's example was solved using Algorithm~\ref{alg:GDrandsearch} with
the exact (Lyapunov-based) calculation for $J_{RL}$ and $\nabla
J_{RL}$.  The implementation used $N_{ri} = 3\times
10^4$ random initializations and $N_{ga} =
100$ gradient steps for each initialization.  The following hypercube
was used for sampling:
\begin{align}
  \underline{\theta} := \bmtx -0.2 \\ -0.2 \\ -40 \\ 0 \emtx
  \mbox{ and }
  \bar{\theta} := \bmtx 0 \\ 0 \\ 0 \\ 40 \emtx
\end{align}



The best policy computed at the end of the search was:
\begin{align}
  \theta_{opt} & := \bmtx -0.0346 & -0.0687 & -20.3441 & 22.83  \emtx^T 
\end{align}
The reward achieved by this sub-optimal policy is
$J_{opt}=-1.488 \times 10^5$.  This is only 8.3\% larger than
the cost achieved by the optimal LQG controller (accounting
for the sign change).  The search dimension $\R^4$ is relatively small
and hence this method finds a nearly optimal controller.

The feedback system of the plant and RL policy has classical gain,
phase, and symmetric disk margins of $[0.9633,1.0109]$, $\pm 0.331$
degs, and $m_d=1.009$. The model for the environment (plant) dynamics
was used to compute these margins. However, it is possible to estimate
robustness margins only from data.  These small margins again indicate
that the feedback system may become unstable due to small changes in
the plant gain or parasitic dynamics. This has practical implications
for model-free RL.  Small robustness margins imply that an RL
controller trained via simulation might lead to an unstable feedback
system when implemented on the real plant.  Alternatively, consider
the scenario where the RL controller is trained via experimental data
on a real system. The same RL controller might cause instability if
the dynamics of the system vary slightly over time.  Moreover, the
same RL controller might cause instability if implemented for
production on many devices of the same type (e.g. RL trained on one
robot but implemented for production on many of the same type of
robot).  In summary, LQG is a special case of RL and hence it follows
that RL with POMDPs can also have poor robustness margins.


\subsection{Recovering Robustness in RL}
\label{subsec:recoverRL}

As noted above, typical algorithms to improve robustness are model
based. It would be useful to have an easily implementable, data-driven
method to recover robustness.  Two options for enhancing robustness
are to: (i) alter the POMDP dynamics used in the training process or
(ii) modify the reward function.  This section focuses on the first
option but concludes with a brief comment on the second option.

\begin{figure}[h]
\centering
\scalebox{0.9}{
\begin{picture}(205,135)(-50,-70)
 \thicklines
 \put(7,63){$w$}
 \put(10,60){\line(0,-1){25}}  
 \put(10,35){\vector(1,0){30}}  
 \put(40,0){\framebox(50,50){System}}
 \put(90,25){\vector(1,0){27}}  
 \put(120,25){\circle{6}}  
 \put(117,63){$v$}
 \put(120,60){\vector(0,-1){32}}  
 \put(123,25){\line(1,0){20}}  
 \put(143,25){\line(0,-1){70}}  
 \put(147,-5){$y$}
 \put(143,-45){\vector(-1,0){53}}  
 \put(40,-70){\framebox(50,50){Control}}
 \put(40,-45){\line(-1,0){90}}  
 \put(-50,-45){\line(0,1){60}} 
 \put(-50,15){\vector(1,0){20}}   
 \put(-30,0){\framebox(30,30){$\Delta$}}
 \put(0,15){\vector(1,0){40}}  
 \put(15,5){$u$}
\end{picture}
} 
\caption{RL Training With Input Perturbations}
\label{fig:RLwithIPert}
\end{figure}
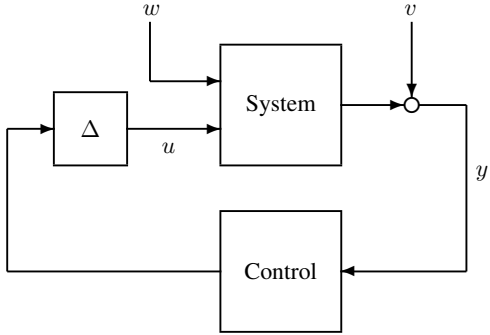

Consider the feedback interconnection shown in
Figure~\ref{fig:RLwithIPert}.  This diagram shows a system
(environment) in feedback with a controller (policy).  The system is
assumed to be modeled by a POMDP with process noise $w$ and sensor
noise $v$ (or more generally by the state transition $\mathcal{T}$ and
observation $\mathcal{O}$ probabilities). The additional box $\Delta$
will be used to introduce perturbations to the dynamics (model
uncertainty) during the training phase.  Temporarily assume that
$\Delta = 1$, i.e. no model uncertainty.  A standard RL training
approach would evaluate the expected reward for the policy over the
random process and sensor noise. One might conjecture that robustness
would be enhanced by increasing the process noise $w$ during the
training phase. In fact, the robustness margins become smaller for
Doyle's example as the process noise variance $W \rightarrow
\infty$. This counterintuitive result emphasizes the distinction
between process noise (which enters externally to the feedback system)
and model uncertainty (which appears internally in the feedback
system).

The proposed method to enhance robustness is to perform the training
with random input perturbations.  If the input is scalar then the
perturbation in Figure~\ref{fig:RLwithIPert} is set as
$\Delta=1+\delta$ where $\delta$ is a uniform random variable in
$[-b,b]$.\footnote{Another alternative is to set $\Delta$ as a uniform
  random variable in $[\frac{1}{m}, m]$ where $m>1$.  This would align
  with the disk margin definition.}  The value $b>0$ is selected to
tune the amount of desired margin.  The expected reward is computed
over these input perturbations as well as the process and sensor
noise. Thus the perturbations should be randomly sampled during each
data collection.  Such perturbations can be easily introduced during
model-free RL training.  They can be introduced when training either
with simulations or with experimental devices.  If the system has
multiple inputs then a similar (independent) perturbation can be
introduced into each input channel. The perturbation level $b_i$ can
be specified uniquely for channel $i$ to obtain a desired robustness
margin for that channel.

\begin{figure}[h]
\centering
\includegraphics[width=\linewidth]{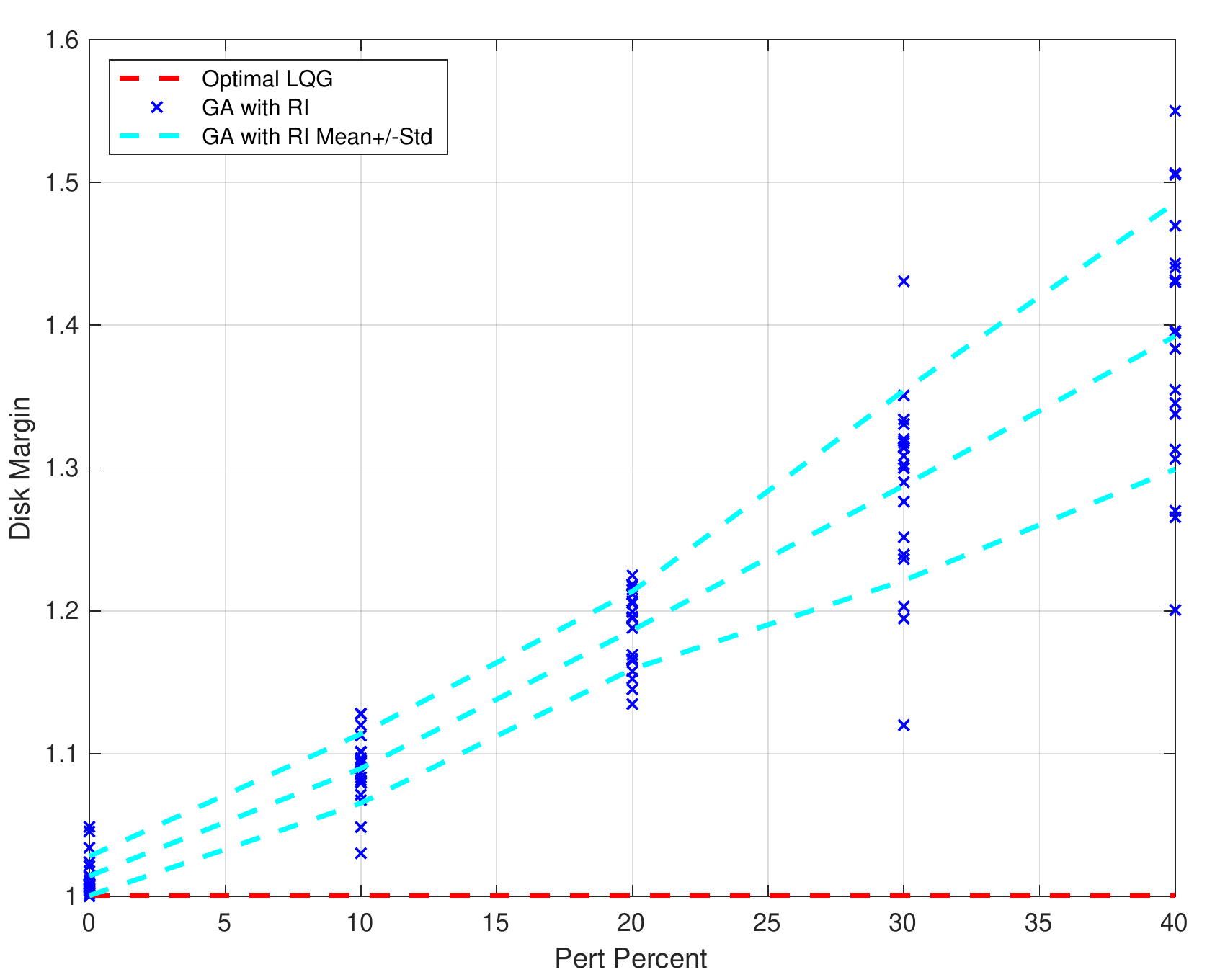}
\caption{Disk Margin vs Input Perturbation Percent ($100 \times b$)}
\label{fig:DoyleIPdm}
\end{figure}

\begin{figure}[h]
\centering
\includegraphics[width=\linewidth]{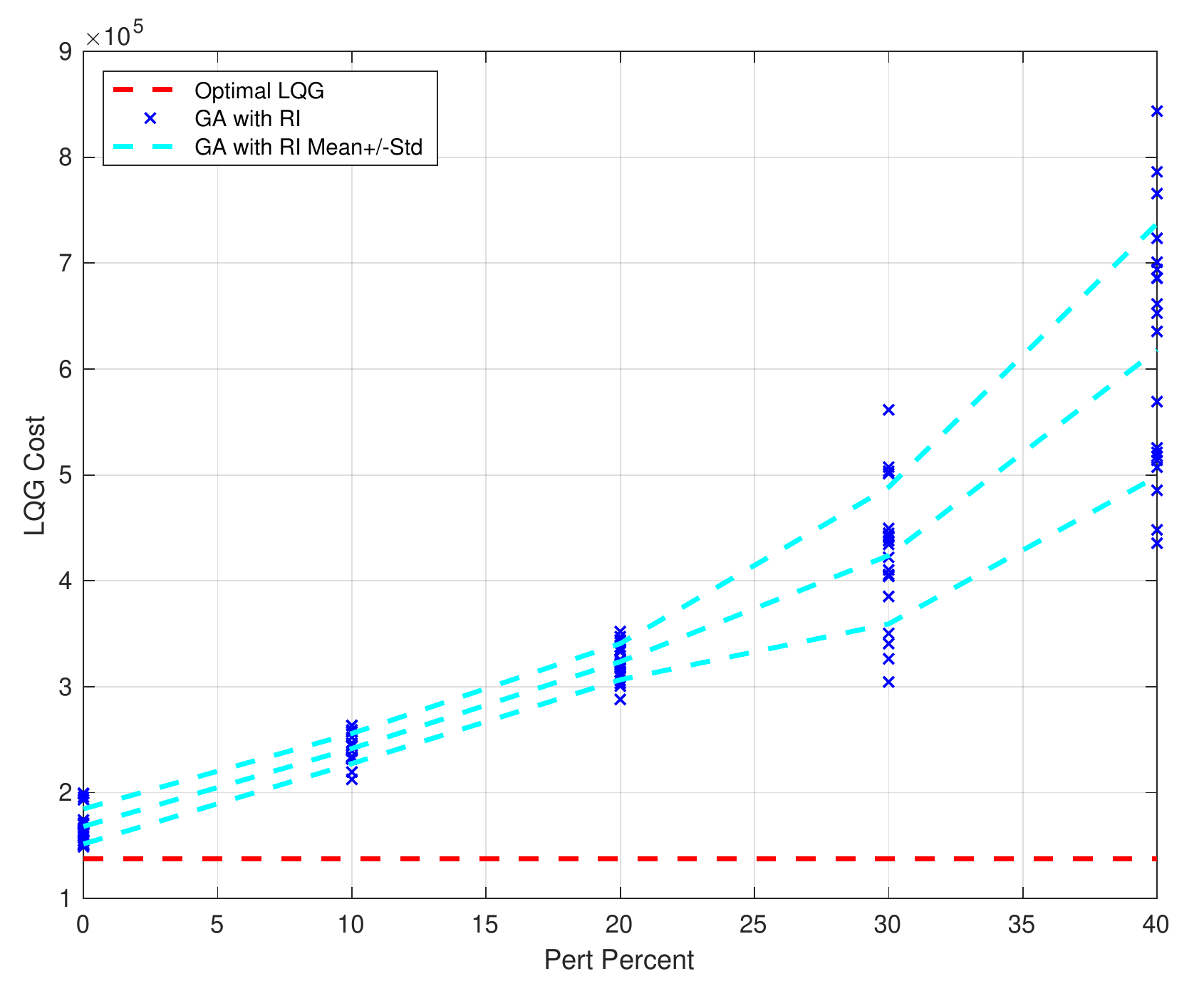}
\caption{LQG Cost vs Input Perturbation Percent ($100\times b$)}
\label{fig:DoyleIPcost}
\end{figure}

Gradient ascent was again applied to Doyle's example with input
perturbations at levels $b=0$ (No perturbation), 0.1, 0.2, 0.3, 0.4.
The algorithm parameters $(\mathcal{H},N_{ri},N_{ga},\eta)$ were
chosen the same as in the previous section.
Figure~\ref{fig:DoyleIPdm} shows the disk margin versus the input
perturbation percentage $(100 \times b) \%$.  The algorithm was
repeated 20 times for each input perturbation level. Each blue
\texttt{x} corresponds to one of these trials. The mean and $\pm$ one
standard deviation of these trial results are shown as cyan dashed
lines.  Finally, the disk margin for the optimal LQG controller
$m_d=1.0007$ is shown as the flat dashed red line.  The disk margin
increases with the input perturbation level
$b$. Figure~\ref{fig:DoyleIPcost} shows the corresponding LQG costs
(equal to the negative of the expected reward) versus the perturbation
percent.  This figure also shows the results for each of the one
twenty trials (blue \texttt{x}), mean and $\pm$ one standard deviation
over all trials (dashed cyan), and optimal LQG cost (red dashed at
$J =1.373 \times 10^5$).  The cost increases (decreasing reward) as
the input perturbation increases.  This shows the expected robustness
versus performance trade-off.  The perturbation level $b$ provides a
``knob'' to easily make this trade-off.

To conclude this section, we briefly comment on the option to enhance
robustness, in the RL framework, through proper modification of the
reward function $r$. Consider the following per timestep reward for
Doyle's example:
\begin{align}
  r_{LQG}(x_t,u_t) := -\left( \sigma \, 
          x_t^T \bsmtx 1 & 1 \\ 1 & 1 \esmtx x_t + u_t^T u_t \right)
\end{align}
If $\sigma = 1000$ then this corresponds to the reward used in the
previous section to solve Doyle's example in RL.  This yielded small
robustness margins.  The margins become progressively smaller for
$\sigma \rightarrow \infty$ as noted in \cite{doyle78}. Thus
increasing the state penalty (or decreasing the control effort
penalty) will further degrade robustness on this example.  Conversely,
robustness is enhanced on Doyle's example by reducing the reward for
good disturbance rejection.  Trading performance vs. robustness via
properly modifying the reward function can be difficult or
counter-intuitive in more complex problems.  The input
perturbation method described above provides a more direct means to
improve robustness.



\section{Example: Simplified Flexible System}
\label{sec:application}

This section considers a simplified flexible aircraft model drawn from
\cite{moore81,anderson07}. The model given in \cite{anderson07} is in
continuous-time and represents a system with at low frequency rigid
body mode at 1 rad/sec and a lightly damped flexible mode at
10rad/sec.  The model also includes a coloring filter on the process
noise with a bandwidth at 1 rad/sec.  This idealized model can
represent any system with a dominant low frequency rigid motion and
high frequency flexible mode, e.g. a robotic system.

The continuous-time model was discretized with a zero-order hold and
sample time $T_s=0.09$sec.  The corresponding plant, noise, and cost
data for a discrete-time LQG problem (as formulated in
Section~\ref{subsec:lqg}) is given by:
\begin{align*}
A & := \bmtx 0.9139 & 0 & 0 & 0.0823 \\
        0 & 0.6238 & 0.0776 & 0  \\
        0 & -7.7632 & 0.6083 & 0 \\
        0 & 0 & 0 & 0.9139 \emtx \\ 
B &:= \bmtx 0.0861 \\ 0.3762\\ 7.7632\\ 0 \emtx, 
B_w := \bmtx 0.0017 \\ 0 \\ 0 \\0.0387 \emtx,
C^T := \bmtx 1 \\ 10 \\ 0 \\ 1 \emtx \\
Q &:= \bmtx 4 & 0& 0& 0 \\ 0 &  0& 0& 0  \\ 
      0 &  0& 0& 0  \\ 0 &  0& 0& 0\emtx, 
R := 1, W:=1, V := 0.01
\end{align*}

The optimal LQG controller is the estimator and state
feedback in Equation~\ref{eq:lqg} with the following gains:
\begin{align*}
   K  & =  \bmtx 1.1154 & 0 & 0 & 0.3976 \emtx \\ 
   L^T & = \bmtx 0.0673 & 0 &  0 & 0.2496 \emtx
\end{align*}
The optimal cost achieved by this LQG controller is $J_{LQG} = 0.0072$
and the disk gain margin is $m_d = 1.0091$.  This small margin again
indicates the poor robustness of the optimal (model-based) LQG
controller for this system.

Standard RL can be used to construct controllers for this example as
discussed in Section~\ref{subsec:solvinglqg}.  The policy is
parameterized as a third-order system (as in Eq.~\ref{eq:lqgcanon})
with state matrices in controllable canonical form:
\begin{align*}
  A_K(\theta) := \bmtx 0 & 1 & 0 \\ 0 & 0 & 1  \\ 
                     \theta_1 & \theta_2 & \theta_3 \emtx, 
  B_K(\theta) := \bmtx 0 \\ 0 \\ 1 \emtx,
  C_K^T(\theta) := \bmtx \theta_4 \\ \theta_5 \\ \theta_6 \emtx                      
\end{align*}
Each vector $\theta \in \R^6$ corresponds to a specific policy denoted
by $K(\theta)$.  The gradient ascent in
Algorithm~\ref{alg:GDrandsearch} is applied with $N_{ri} = 10^4$ random
initializations and
$N_{ga} = 10^3$ gradient steps. The search hypercube
$\mathcal{H}( \underline{\theta} , \bar{\theta} )$ is defined by:
\begin{align*}
  \underline{\theta} & := \bmtx 0 & -2 & 0 & -0.1 & 0 & -0.3 \emtx^T, 
  \\
  \bar{\theta} & := \bmtx 1 & 0 & 2 & 0 & 0.3 & 0  \emtx^T
\end{align*}
These bounds were chosen with some trial and error. Most controllers
in this search space are stable and minimum phase.   In practice,
some a priori knowledge would be required to obtain reasonable
bounds on the search space.

Algorithm~\ref{alg:GDrandsearch} was repeated for 25 trials with no
input perturbations. The optimal controller found on these trials
achieved a cost between 0.0075 and 0.0085. Thus the RL implementation
converges to nearly optimal controllers.  Small disk margins were
obtained for these RL controllers with values ranging from 1.0131 to
1.31.  Note that the optimal LQG controller is fourth-order and is not
contained within the third-order parametrization used for RL.  These
results demonstrate that robustness issues in RL can still arise even
with parameterizations that do not include the optimal LQG
controller. This further motivates the need to recover robustness in
the RL training.

\begin{figure}[ht]
\centering
\includegraphics[width=\linewidth]{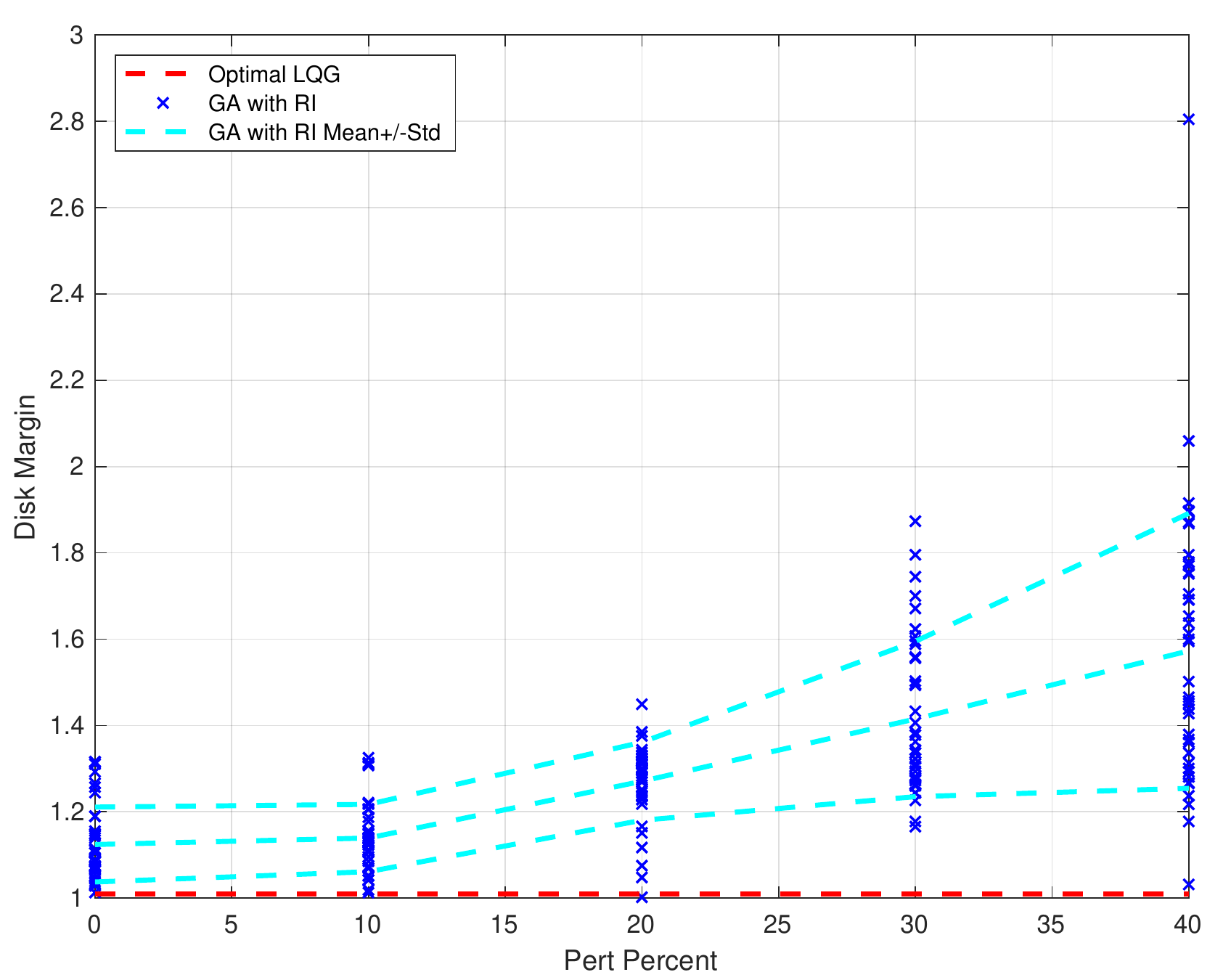}
\caption{Disk Margin vs Input Perturbation Percent ($100 \times b$)}
\label{fig:MooreIPdm}
\end{figure}
\begin{figure}[ht]
\centering
\includegraphics[width=\linewidth]{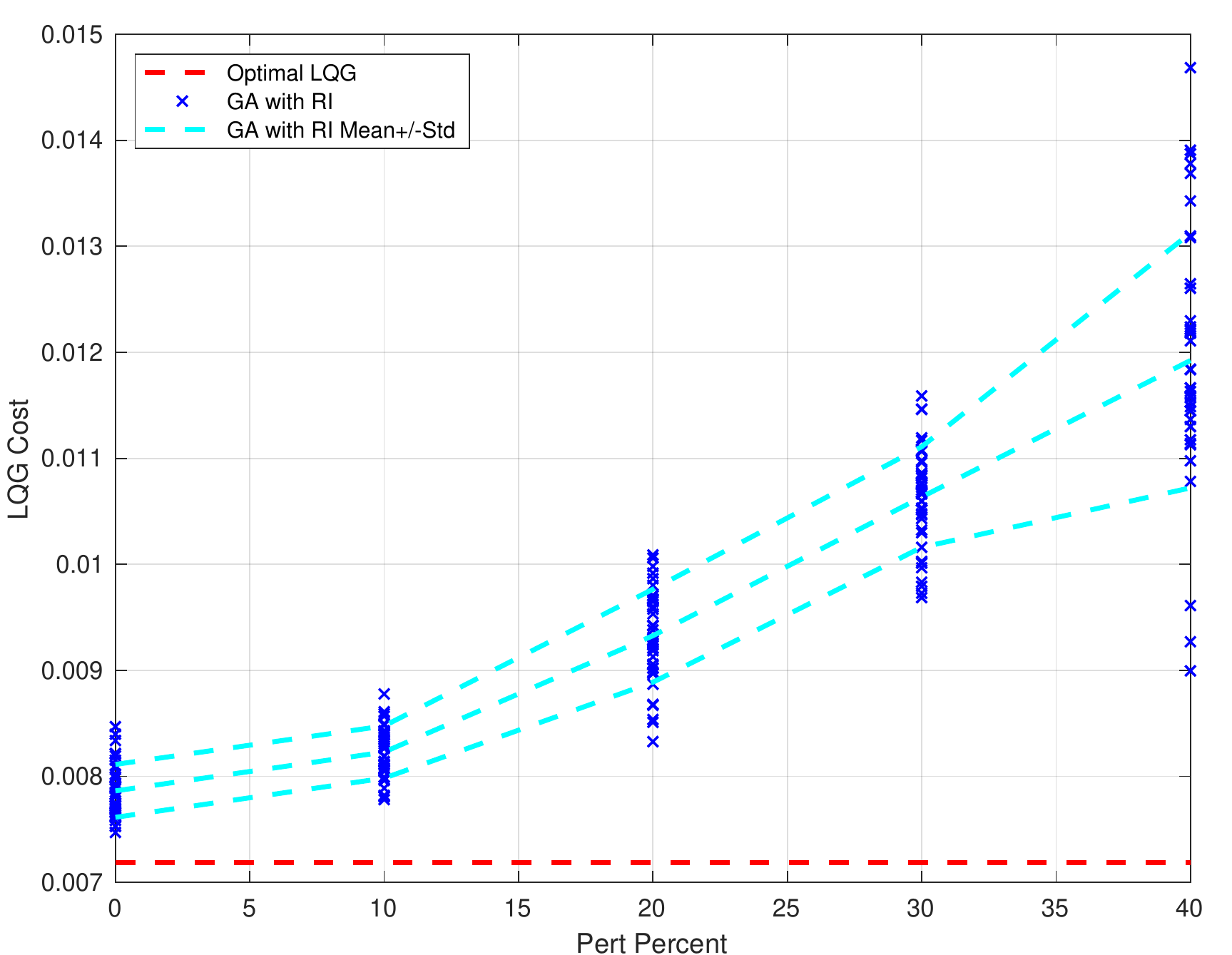}
\caption{LQG Cost vs Input Perturbation Percent ($100 \times b$)}
\label{fig:MooreIPcost}
\end{figure}

The input perturbation method (Section~\ref{subsec:recoverRL}) was
applied to this example with perturbation levels $b=0$ (No
perturbation), 0.1, 0.2, 0.3, 0.4.  Figures~\ref{fig:MooreIPdm} and
\ref{fig:MooreIPcost} show the disk margins and LQG cost ($=-$reward)
for this example. The input perturbation method was applied with the
same $N_{ri}$, $N_{ga}$ and hypercube
$\mathcal{H}( \underline{\theta} , \bar{\theta} )$ specified above.
These figure shows the results for each of the twenty five trials
(blue \texttt{x}), mean and $\pm$ one standard deviation over all
trials (dashed cyan), and for the optimal LQG controller (flat red
dashed).  Figure~\ref{fig:MooreIPdm} shows the improvement in the disk
margin robustness with increasing input perturbation level.
Conversely, Figure~\ref{fig:MooreIPcost} shows the degradation in
performance with increasing input perturbation level.  This again
demonstrates that the perturbation level $b$ can be used to trade off
robustness and performance during the RL training.

\section{Conclusion}

Reinforcement learning (RL) with POMDPs is sufficiently general to
solve the standard LQG problem. Thus LQG can be used to explore the
robustness of RL controllers. A simple example by Doyle was used to
demonstrate that RL with partial observations can lead to poor
robustness margins. It is proposed to recover robustness by
introducing random perturbations at the system input during the RL
training.  Two simple examples were used to demonstrate the
effectiveness of this technique to trade off performance for
robustness. Future work will explore the theoretical basis for the
numerical results observed in this paper.



\bibliographystyle{plain}
\bibliography{RLRobustness}

\begin{thebibliography}{10}

\bibitem{anderson07}
B.D.O. Anderson and J~.B. Moore.
\newblock {\em Optimal Control: Linear Quadratic Methods}.
\newblock Dover, 2007.

\bibitem{bates02}
D.~Bates and I.~Postlethwaite.
\newblock {\em Robust multivariable control of aerospace systems}.
\newblock Delft University Press, 2002.

\bibitem{blight94}
J.~D. Blight, R.~L. Dailey, and D.~Gangsaas.
\newblock Practical control law design for aircraft using multivariable
  techniques.
\newblock {\em International Journal of Control}, 59(1):93--137, 1994.

\bibitem{dean17}
S.~Dean, H.~Mania, N.~Matni, B.~Recht, and S.~Tu.
\newblock On the sample complexity of the linear quadratic regulator.
\newblock {\em arXiv}, 2017.

\bibitem{doyle78}
J.~Doyle.
\newblock Guaranteed margins for lqg regulators, in ieee transactions on
  automatic control.
\newblock {\em IEEE Transactions on Automatic Control}, 23(4):756--757, 1978.

\bibitem{doyle82}
J.~Doyle.
\newblock Analysis of feedback system with structured uncertainties.
\newblock {\em IEE Proceedings D - Control Theory and Applications}, pages
  242--250, 1982.

\bibitem{dgkf89}
J.C. Doyle, K.~Glover, P.P. Khargonekar, and B.A. Francis.
\newblock State-space solutions to standard {$H_2$} and {$H_\infty$} control
  problems.
\newblock {\em IEEE Trans. on Aut. Control}, 34(8):831--847, 1989.

\bibitem{doyle79}
J.C. Doyle and G.~Stein.
\newblock Robustness with observers.
\newblock {\em IEEE Trans. on Aut. Control}, 24(4):607--611, 1979.

\bibitem{doyle81}
J.C. Doyle and G.~Stein.
\newblock Multivariable feedback design: Concepts for a classical/modern
  synthesis.
\newblock {\em IEEE Trans. on Aut. Control}, 26(1):4--16, 1981.

\bibitem{gudimella17}
A.~Gudimella, R.~Story, M.~Shaker, R.~Kong, M.~Brown, V.~Shnayder, and
  M.~Campos.
\newblock Deep reinforcement learning for dexterous manipulation with concept
  networks.
\newblock {\em arXiv}, 2017.

\bibitem{kalashnikov18}
D.~Kalashnikov, A.~Irpan, P.~Pastor, J.~Ibarz, A.~Herzog, E.~Jang, D.~Quillen,
  E.~Holly, M.~Kalakrishnan, V.~Vanhoucke, and S.~Levine.
\newblock {QT-Opt}: Scalable deep reinforcement learning for vision-based
  robotic manipulation.
\newblock {\em arXiv}, 2018.

\bibitem{kwak72}
H.~Kwakernaak and R.~Sivan.
\newblock {\em Linear Optimal Control Systems}.
\newblock Wiley, 1972.

\bibitem{moore81}
J.B. Moore, D.~Gangsaas, and J.D. Blight.
\newblock Performance and robustness trades in lqg regulator design.
\newblock In {\em IEEE Conference on Decision and Control}, pages 1191--1200,
  1981.

\bibitem{ng04}
A.Y. Ng, H.~J. Kim, M.~I. Jordan, and S.~Sastry.
\newblock Autonomous helicopter flight via reinforcement learning.
\newblock In {\em Advances in Neural Information Processing Systems 16}, pages
  799--806, 2004.

\bibitem{packard93mu}
A.~Packard and J.~Doyle.
\newblock The complex structured singular value.
\newblock {\em Automatica}, 29(1):71--109, 1993.

\bibitem{packard93dk}
A.~Packard, J.~Doyle, and G.~Balas.
\newblock Linear, multivariable robust control with a $\mu$ perspective.
\newblock {\em Transactions of the {ASME}}, 115:426--438, 1993.

\bibitem{paganini95}
F.~Paganini.
\newblock Robust {$H_2$} performance: Guaranteeing margins for lqg regulators.
\newblock Technical Report CIT-CDS 95-031, CDS at California Institute of
  Technology, 1995.

\bibitem{peters06}
J.~Peters and S.~Schaal.
\newblock Policy gradient methods for robotics.
\newblock In {\em {IEEE/RSJ} International Conference on Intelligent Robots and
  Systems}, pages 2219 -- 2225, 2006.

\bibitem{recht18}
B.~Recht.
\newblock A tour of reinforcement learning: The view from continuous control.
\newblock {\em arXiv}, 2018.

\bibitem{stulp10}
F.~Stulp, J.~Buchli, E.~Theodorou, and S.~Schaal.
\newblock Reinforcement learning of full-body humanoid motor skills.
\newblock In {\em {IEEE-RAS} International Conference on Humanoid Robots},
  pages 405 -- 410, 2010.

\bibitem{sutton98}
R.S. Sutton and A.G. Barto.
\newblock {\em Reinforcement Learning: An Introduction}.
\newblock Bradford, 1998.

\bibitem{szepesvari10}
C.~Szepesvari.
\newblock {\em Algorithms for Reinforcement Learning}.
\newblock Morgan and Claypool, 2010.

\bibitem{zhou95}
K.~Zhou, J.C. Doyle, and K.~Glover.
\newblock {\em Robust and Optimal Control}.
\newblock Pearson, 1995.

\end{thebibliography}

\appendix

\label{app:ssvar}


Consider the following discrete-time system:
\begin{align}
\label{eq:GenLTI}
  \bar{x}_{t+1} & = \bar{A} \bar{x}_t + \bar{w}_t 
\end{align}
where $\bar{w}_t$ is white, zero mean, and Gaussian with variance
$\bar{W}:=E[\bar{w}_t \bar{w}_t^T]$.  Assume $\bar{A}$ is a Schur
matrix, i.e. all eigenvalues have magnitude $<1$. There exists a
unique solution $X\succeq 0$ to the discrete-time Lyapunov equation:
\begin{align}
  \bar{A}^TX\bar{A}-X+\bar{W} = 0
\end{align}
The following steady-state relation holds for any matrix $M$:
\begin{align}
\label{eq:SSCost}
  \lim_{N\rightarrow \infty} \frac{1}{N} 
  E \left[ \sum_{t=0}^N \bar{x}_t^T M \bar{x}_t \right]
  = trace\left( MX \right)
\end{align}

The dynamics of the plant (environment) and controller (policy) can be
combined to model the closed-loop system as in
Equation~\ref{eq:GenLTI}. Moreover, the RL cumulative reward can be
expressed as in Equation~\ref{eq:SSCost} for an appropriately chosen
$M$. Thus this result can be used to exactly compute the closed-loop reward
from the solution of the Lyapunov equation.

The gradient of the expected cumulative reward function can  be
obtained by evaluating the gradient of Equation~\ref{eq:SSCost}.
An application of the chain rule yields:
\begin{align}
\label{eq:SSCostGrad}
\nabla J_{RL}(\theta) = trace\left(\nabla M(\theta) \cdot \, X(\theta) 
         + M(\theta) \cdot \, \nabla X(\theta) \right)  
\end{align}
Assume that both $\nabla
M(\theta)$ and $\nabla \bar{A}(\theta)$ can be computed. Then
only $\nabla X(\theta)$ is needed in order to compute $\nabla J_{RL}(\theta)$
This is obtained by solving the following Lyapunov equation
for $\nabla X$ (dropping the notational dependence on $\theta$):
\begin{align}
\bar{A}^T \, (\nabla X) \,\bar{A}
- \nabla X + 
(\nabla\bar{A})^T \, X \, \bar{A}
+\bar{A}^T \, X\, (\nabla\bar{A}) = 0   
\end{align}


\end{document}